# Single-Pulse Optical Switching combined with Current-Induced Motion of Skyrmionic Spin Textures

Edoardo Mangini[1], Duc Minh Tran[1], Boonthum Kunyangyuen[2], Jun Xiao Lin[2], Raphael Gruber[1], Michel Hehn[2,3], Stéphane Mangin*[2,3], and Mathias Kläui*[1]

[1]Institut für Physik, Staudingerweg. 7, 55128, Mainz, Germany, Email Address: klaeui@uni-mainz.de
[2]Institut Jean Lamour, Université de Lorraine, CNRS Nancy, France, Email Address: stephane.mangin@univ-lorraine.fr
[3]Center for Science and Innovation in Spintronics, Tohoku University, Sendai, Japan

**Abstract**

Magnetic skyrmions are promising nanoscale information carriers because their position can be manipulated by electrical currents. The additional ability of deterministic control of the spin structure would allow the use of magnetic skyrmions as multi-state bits, expanding the capabilities of skyrmion devices further. Such control could be realized by coupling all-optical helicity independent switching of magnetization (AO-HIS) to magnetic skyrmions. This is realized in a Pt/Ir/CoB/Gd/Pt multilayer that is engineered to stabilize magnetic skyrmions and, at the same time, enable AO-HIS by single ultrafast laser pulse. Laser excitation is used to write skyrmionic textures, whose polarity is selected by a small applied out-of-plane field. A single 30 fs laser pulse is then used to toggle the magnetization of an illuminated region containing skyrmionic textures, reversing their polarities without requiring any magnetic field or current pulses. This establishes deterministic optical switching of the skyrmion spin structure as a new manipulation channel, distinct from previously reported optical nucleation and annihilation. The material is then patterned into a wire, to combine the optical manipulation with current-induced skyrmion motion. These results establish a route toward opto-spintronic skyrmion devices in which light programs the internal state of a skyrmion and electrical currents control its position.

## 1 Introduction

Magnetic skyrmions are chiral, topologically non-trivial spin textures that can be stabilized in magnetic multilayers with perpendicular magnetic anisotropy and interfacial Dzyaloshinskii-Moriya interaction (DMI) [1–3]. Their small size, robustness and sensitive response to external stimuli have made them attractive information carriers for low-power spintronic applications, including racetrack memories [3–6], sensors [7, 8], stochastic and Brownian reservoir computing [9, 10], and neuromorphic devices [11]. In these concepts, the ability to create, erase, displace and detect skyrmions is central to device operation.

Current-induced manipulation is an established route to skyrmion control. Spin-orbit torques (SOT) generated by current injection in heavy-metal/ferromagnet stacks can drive skyrmions deterministically along patterned tracks [5, 12, 13]. Additionally, magnetic fields and field gradients [14, 15], temperature [16, 17], device geometry [10, 16], and electric gating [18] can tune their stability, density and dynamics. These mechanisms provide wide possibilities for the design of skyrmion-based devices, where skyrmions are deterministically nucleated, annihilated, and displaced. Deterministic control over the spin structure of a magnetic skyrmion is however still a key missing piece, which can expand the capabilities of skyrmion based devices. As an example, in a device where the spin structure of a magnetic skyrmion can be toggled between multiple states, the spin state of the skyrmion can be used to store information.

Optical excitation has been recently used to manipulate the spin structure of magnetic textures. Ultrafast laser pulses can heat magnetic materials on femtosecond timescales and trigger magnetization reversal in ferrimagnetic alloys and multilayers through all-optical helicity-independent switching (AO-HIS) [19–23]. In Gd-Co alloys and multilayers, time-resolved measurement techniques have shown that the phenomenon can occur on ultra-fast, sub-picosecond timescales [20, 24]. Optical excitation has also been used to nucleate skyrmions stochastically or deterministically [25, 26], and recent work has shown that single ultrafast laser pulses can write and annihilate skyrmions by AO-HIS [27]. These demonstrations establish optical writing and erasing of magnetic domains as powerful tools. However, they still do not provide deterministic control of the spin structure while preserving the magnetic texture.

AO-HIS can also be used to manipulate the spin structure of a skyrmion, without annihilation, by inverting the sign of the magnetization of an area containing the skyrmion. When this operation is performed, the skyrmion core polarity (the out of plane component of the core magnetization) is also reversed, and a more complex structure is obtained. One example of such a structure is a concentric target skyrmion [28, 29]. By switching back the magnetization of the area, the initial skyrmion state is recovered. Thus, the control over skyrmion polarity via AO-HIS enables the use of multi-state skyrmion bits, which can store additional information in their internal state, and not only in their position or presence. These new information carriers can be exploited in optically programmable racetracks, local reset operations, and hybrid photonic-spintronic schemes, where light toggles the internal state while electrical currents transport the information by moving the skyrmionic information carrier.

In this work, we demonstrate AO-HIS of magnetic skyrmion textures, and combine this functionality with current-driven skyrmion motion in the same material. We design a Pt/Ir/CoB/Gd/Pt multilayer where the Pt/Ir/CoB structure provides perpendicular magnetic anisotropy, interfacial DMI and spin-orbit-torque functionality [30–32], while the CoB/Gd ferrimagnetic system enables AO-HIS [21, 22, 24]. We nucleate magnetic skyrmions in our material via laser excitation, and subsequently exploit AO-HIS to deterministically switch their core polarities using single ultra-fast laser pulses. Finally, we use the material to fabricate a wire device, where we use spin-polarized currents to drive the magnetic textures, demonstrating that our platform enables the coupling between skyrmion core polarity manipulation and current-driven motion.

## 2 Results and discussion

### 2.1 Material platform for skyrmion stability and all-optical switching

The schematics of **Figure 1a-d** show a proposed multi-bit skyrmion device operation. Skyrmions are nucleated via pulse-train laser excitation, the polarity of one skyrmion is switched with a single laser pulse, and the textures are displaced by the application of a spin polarized current. In order to enable such operations, we designed a material that allows both stabilization of skyrmions and AO-HIS. We used a multilayer with composition Ta(3)/Pt(5)/Ir(1)/CoB($t_{CoB}$)/Gd($t_{Gd}$)/Pt(2), where all thicknesses are given in nm, as shown in Figure 1e. The Pt/Ir/CoB interfaces provide perpendicular magnetic anisotropy and interfacial Dzyaloshinskii - Moriya interaction (DMI) [30, 33]. Additionally, the broken inversion symmetry and heavy-metal underlayers enable current-induced skyrmion motion via SOT [32]. The addition of Gd is necessary to enable AO-HIS [21,22,24]. Adding Gd also introduces ferrimagnetic character [34], and tunes the DMI [31]. Because the stabilization of skyrmions and the occurrence of AO-HIS both depend sensitively on layer composition [24, 35, 36], CoB and Gd thicknesses were systematically varied by depositing them as orthogonal wedges on a 4-inch Si/$SiO_2$ wafer, in order to identify regions where both phenomena coexist (see also Supporting Information, **Figure S1**).

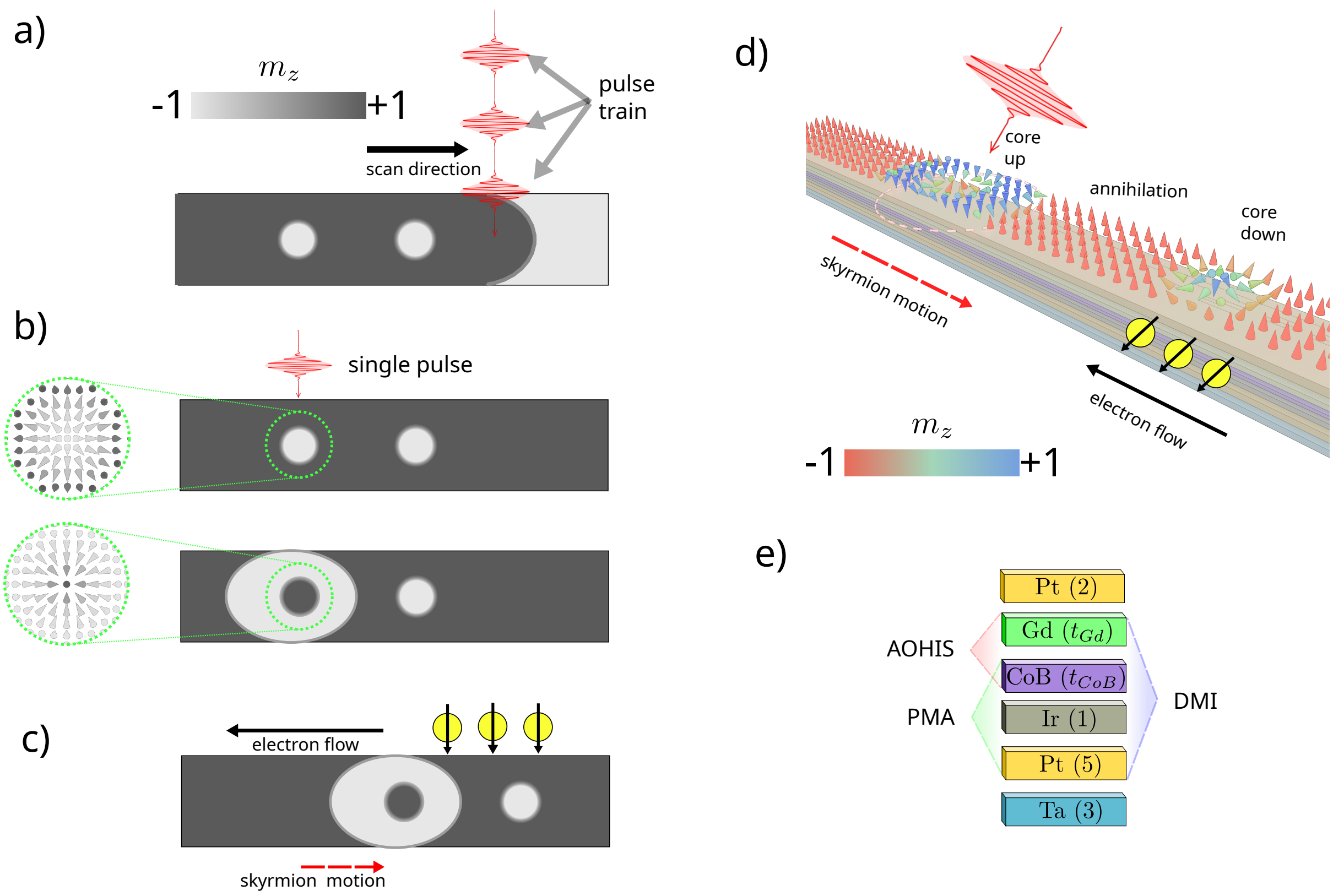

Figure 1: Concept and material platform. a-c) Schematic of the opto-electrical skyrmion manipulation sequence: pulse-train laser excitation writes skyrmions (a), a single femtosecond laser pulse reverses the skyrmion core polarity (b), and current-induced spin-orbit torques transport the skyrmion in a wire (c). d) Device concept illustrating that light pulses program the internal core state while electrical current controls the skyrmion position. e) Multilayer structure Ta/Pt/Ir/CoB/Gd/Pt, where CoB and Gd thicknesses are tuned to combine interfacial DMI, perpendicular magnetic anisotropy and AO-HIS.

We found a composition such that skyrmions can be stabilized, and AO-HIS can be triggered by a single 30 fs linearly polarized laser pulse. This region corresponds to the layer thicknesses of $t_{CoB} = 1.66$ nm and $t_{Gd} = 0.94$ nm. The AO-HIS behaviour of the material is shown in **Figure 2a–c**. A single pulse switches the magnetization of an irradiated region where the local fluence exceeds a threshold value $F_{\text{th}}$. In our experiments, the laser pulse has a Gaussian fluence distribution and circular shape, but the sample is irradiated with an angle of incidence of 50°, resulting in an elliptical shape of the laser beam spot on the sample, and of the switched region. When the average fluence exceeds the value $F_{\text{MD}} = 7.96 \pm 0.59$ mJ cm$^{-2}$, a multidomain state appears at the center of the switched region, and this multidomain switched area increases as fluence is increased. The major and minor semiaxes of the switched domain follow the logarithmic dependence on the average incident fluence $d = D_{x,y}[\log(F/F_{\text{th}})]^{1/2}$, where $d_{x,y}$ is the semiaxis of the switched region, $D_{x,y}$ is the corresponding semiaxis of the laser beam spot, and $F_{\text{th}}$ is the threshold fluence for AO-HIS [37]. By fitting the domain sizes shown in Figure 2d, we extract a single-pulse switching threshold of $F_{\text{th}} = 3.98 \pm 0.29$ mJ cm$^{-2}$. The fluence window which corresponds to deterministic optical switching, without any stochastic generation of magnetic domains, is the interval between $F_{\text{th}}$ and $F_{\text{MD}}$ [19, 21, 37].

## 2.2 Optical writing of skyrmions

After characterizing the AO-HIS behaviour of the material, we next explore skyrmion manipulation. We generate magnetic skyrmions in our material by utilizing the laser in pulse-train mode while scanning the laser across the sample, as shown in **Figure 3**. We use a pulse duration $\tau = 7.5$ ps, at a repetition rate $f = 5$ kHz, with a fluence $F = 13.28$ mJ cm$^{-2}$. In this nucleation method, small magnetic fields, well below the coercivity of the material, can be used to select the nucleated magnetic structures, as shown in **Figure S2**, and the textures are stable after nucleation, even if the field is removed. This can be explained with the transient reduction of domain-wall pinning during laser-induced heating [38]. When the pulse train passes across a given point, the local temperature rises, domain walls become less strongly pinned, and a small out-of-plane magnetic field can bias the final magnetic configuration. When the laser moves away, the local temperature decreases, pinning increases again, and the skyrmions are stabilized against further changes of the magnetic field. Optical writing therefore provides a practical route to tailor and nucleate an initial skyrmion population.

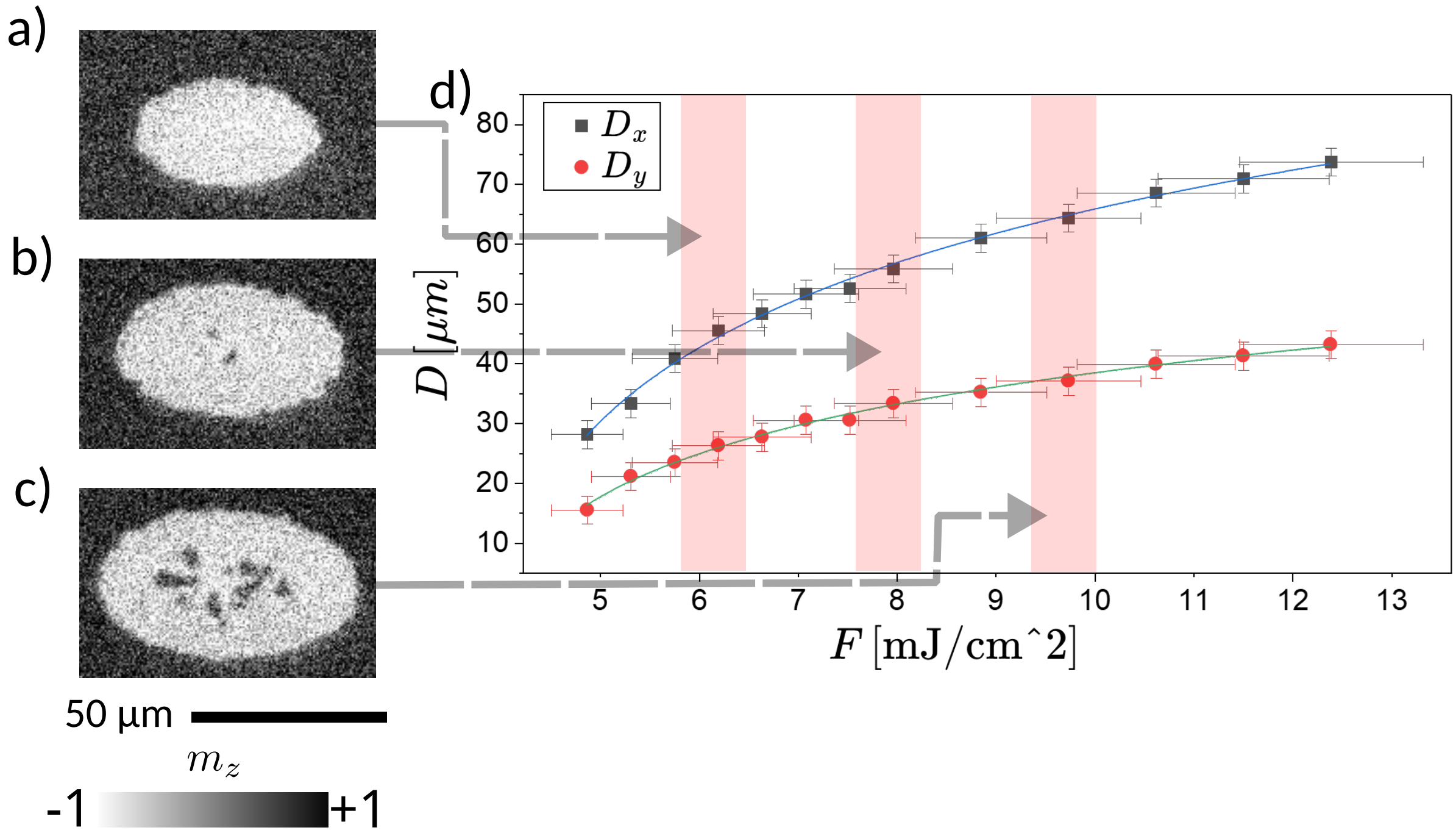

Figure 2: AO-HIS operating window in the CoB/Gd skyrmion platform. a–c) Kerr microscopy images of switched regions obtained at representative fluences, showing deterministic toggle switching below the multidomain threshold and multidomain formation above $F_{\mathrm{MD}} = 7.96 \pm 0.6\ \mathrm{mJ\,cm^{-2}}$. The images correspond to increasing fluence, as indicated by the arrows. d) Size of the single-pulse switched domains as a function of laser fluence for the continuous-film sample.

The major and minor ellipse axes, $D_x$ and $D_y$, are fitted using $d = D[\log(F/F_{\mathrm{th}})]^{1/2}$ [37], yielding an AO-HIS threshold of $F_{\mathrm{th}} = 3.98\ \mathrm{mJ\,cm^{-2}}$.

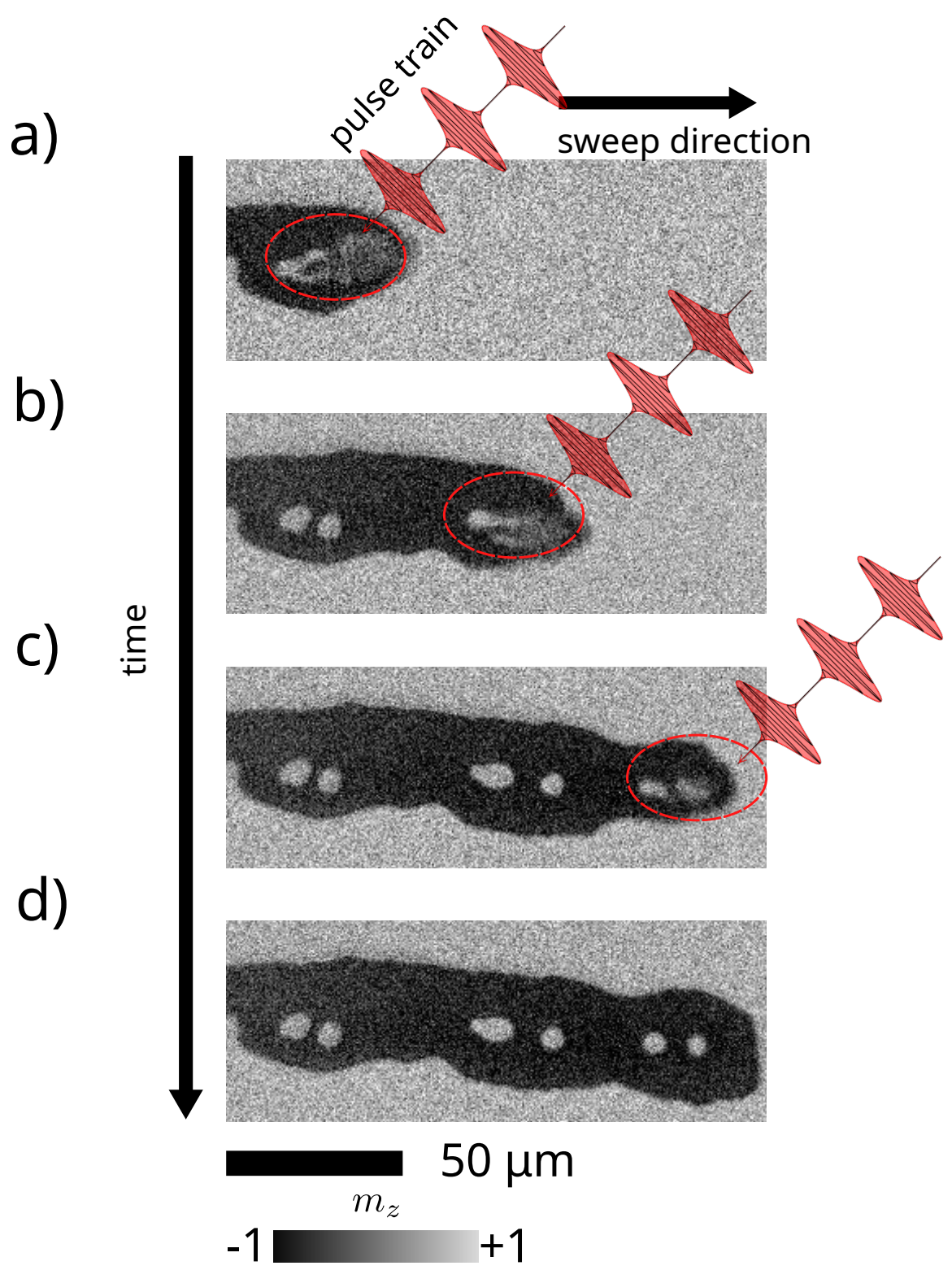

Figure 3: a)-c) Kerr microscopy snapshots of magnetic skyrmions nucleated by pulse-train laser scanning, showing the written skyrmionic textures. Snapshots are taken as the laser spot is swept from left to right, as illustrated by the positions of the dashed ellipses. d) Final state, after nucleation.
Laser parameters: $\tau = 7.5$ ps, $F = 13.28$ mJ cm$^{-2}$, and $f = 5$ kHz. Scale bar: 50 $\mu$m.

### 2.3 Single-pulse optical reversal of the skyrmion core polarity

We investigate the coupling of AO-HIS with magnetic skyrmions in our material. After writing skyrmions with the pulse-train protocol, we use 30 fs single pulses to trigger AO-HIS in regions containing skyrmions. The pulse fluence 7.08 $\mathrm{mJ\,cm^{-2}}$ is chosen within the deterministic AO-HIS window in order to toggle the core polarity of the skyrmions without generating a multidomain state, as explained in Figure 2.

The initial Kerr microscopy image in **Figure 4**a shows two skyrmions written inside a switched magnetic domain, by using the pulse train protocol . We apply a first single laser pulse to the left side of the field of view, containing one skyrmion only. This causes the irradiated area to toggle switch its magnetization, therefore switching the core polarity of the magnetic skyrmion, as shown in Figure 4b. We then apply a second pulse to an area containing both skyrmions, reversing the core polarities of both textures, as shown in Figure 4c. By applying more pulses, we switch the magnetization of the two skyrmions back and forth, without apparent displacement or modification of the magnetic textures, as shown in **Figure S3**.

The observation is distinct from optical skyrmion writing or annihilation. In previous optical schemes, laser excitation was shown to nucleate skyrmions stochastically or deterministically [25, 26], and more recent work demonstrated deterministic optical writing and annihilation of skyrmionic bubbles through AO-HIS [27]. In our experiments, the skyrmion diameter is smaller than the AO-HIS-switched area. In this case, the laser pulse toggles the magnetization of the illuminated region, including both the skyrmion core and its surrounding magnetic background, thus producing a stable skyrmion with the opposite core polarity rather than erasing it. The inner skyrmion-like feature remains visible after excitation, while the magnetization within the surrounding optical switching footprint is reversed. The experiment therefore demonstrates a reproducible optical transformation of a skyrmion-containing magnetic texture.

Several features support the interpretation of the experiment of Figure 4 in terms of AO-HIS. First, the pulse fluence is below the multidomain threshold identified in Figure 2, which excludes thermally induced random domain nucleation as the origin of the reversed contrast. Second, the switched region follows the deterministic AO-HIS footprint expected for single-pulse optical toggle switching [37], as indicated by the dashed ellipses in Figure 4b–c and Figure S3. Third, the reversed textures remain localized and stable after the pulse, demonstrating that the process leads to a well-defined remanent magnetic state. Fourth, multiple skyrmions can be switched simultaneously within the same optical footprint, as seen in Figure 4c, showing that the manipulation is not tied to a particular defect site but instead results from the local optical toggle operation itself. Because the magnetic state is imaged only before and after excitation, the present measurements do not resolve the reversal dynamics.

This experiment expands the capabilities of optical manipulation of magnetic microstructures. The skyrmion core polarity constitutes a degree of freedom of the magnetic texture, which can be accessed in a local and deterministic way through optical excitation. The information state can now be encoded not only in the presence or absence of a skyrmion and in its position, but also in the sign of its core and surrounding magnetization. Optical core switching provides a route to program skyrmion core polarity, complementing electrical transport methods based on current-induced motion [5, 12, 39, 40].

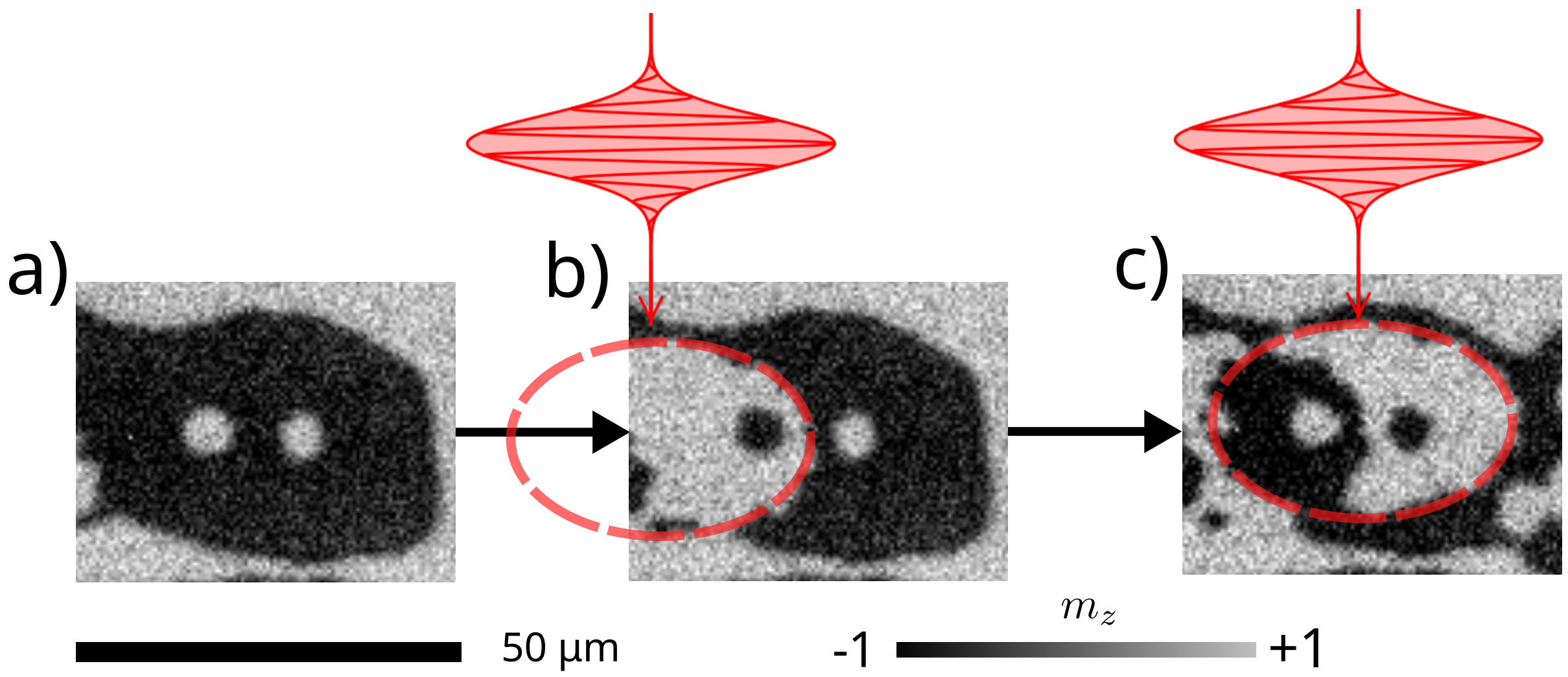

Figure 4: Single-pulse optical reversal of skyrmion core polarity. a) Initial Kerr microscopy image of laser-written skyrmions in the continuous-film sample. b) Image acquired after one 30 fs laser pulse applied to the dashed elliptical region, showing reversal of the selected skyrmion core. c) Image acquired after a second pulse applied to a central region, showing simultaneous reversal of two skyrmion cores. The dashed ellipses indicate the approximate optical switching footprint corresponding to the AO-HIS threshold. A more accurate evaluation of the irradiated areas can be found in Figure S3.
Laser parameters used for the switching sequence are $\tau = 30$ fs and $F = 7.08$ mJ cm$^{-2}$. Scale bars: 50 $\mu$m.

### 2.4 Current-driven transport

In order to allow the device operation shown in Figure 1a-d, optical control of the skyrmion core must be compatible with electrically induced transport. In order to demonstrate that our material platform allows joint electrical and optical manipulation, we patterned the film into 20 $\mu$m wide wires, enabling current injection. The wire sample has $t_{CoB} = 1.66$ nm and $t_{Gd} = 1.04$ nm, selected to preserve the coexistence of skyrmion stability and optical switching. This device geometry allows us to test whether the optical manipulation functions established in the continuous film can be combined with current-induced skyrmion motion in a technologically relevant structure [39, 40].

We nucleate magnetic skyrmions in our wire device, and apply a DC current with density $J_{\mathrm{DC}} = 5.9 \times 10^9$ A m$^{-2}$ supported by OOP oscillatory field and a temperature T= 350 K (see methods for details). We observe skyrmion motion along the wire, as shown in **Figure 5**a-b. We repeat the experiment with a current of the opposite direction, and this reverses the direction of motion. This is consistent with the expected spin-orbit-torque-driven dynamics of interfacial Néel skyrmions [5, 12, 39, 40].

The same wire device also supports optical skyrmion switching at room temperature. As shown in Figure 5c, in the same device, skyrmions can be first written by pulse-train laser scanning at low fluence and then switched with single 30 fs pulses at higher fluence. The writing step was performed with $\tau = 30$ fs, $F = 1.95$ mJ cm$^{-2}$, repetition rate $f = 5$ kHz, and an applied out-of-plane field $B_{\mathrm{DC}} = 0.15$ mT, whereas the subsequent AO-HIS was achieved with single 30 fs pulses at $F = 2.92$ mJ cm$^{-2}$.

Based on these results, we conclude that optical skyrmion core polarity control and current-driven motion are compatible within the same device platform. Figure 5d shows a possible protocol for device operation based on the functionalities we have demonstrated: skyrmions are nucleated via pulse-train excitation at T=300 K in constant applied OOP field, then single pulse excitation is used to toggle the magnetic textures via AO-HIS, and finally the textures are moved via SOT, at T=350 K.

# 3 Conclusion

In conclusion, we engineered a Pt/Ir/CoB/Gd/Pt multilayer as a new opto-electronic platform to couple all-optical helicity-independent switching to interfacial Néel skyrmions, and allowing current-induced skyrmion motion. We used pulse-train laser excitation to write skyrmions, and single 30 fs pulses to reverse the core polarity of skyrmions. In a patterned wire made of the same material, we additionally showed SOT-driven skyrmion motion. These results establish the skyrmion core polarity as an optically programmable internal degree of freedom and open a path toward opto-spintronic skyrmion devices where light programs the information state and electrical currents move it.

The combination of operations demonstrated here unlocks a new role of light in skyrmionics. Light is not only a heat source for stochastic nucleation or deletion; it can be used as a local gate that programs the magnetic state of a skyrmion. Electrical currents, in contrast, provide the established method to displace skyrmions over a distance. The two control channels are therefore complementary: light addresses the core polarity, while current controls the spatial coordinate.

This separation of functions opens several device directions. In racetrack-like geometries, the skyrmion position could define the memory address while the core polarity encodes an additional bit. In multi-state devices, the states could correspond to no skyrmion, core-up skyrmion and core-down skyrmion (Figure 1d). In neuromorphic or reservoir architectures, femtosecond optical pulses could reconfigure the internal magnetic state of a skyrmion ensemble, while electrical currents provide controlled displacement and interaction [41]. Finally, the compatibility with patterned wires shows that the approach can be integrated into hybrid photonic-spintronic circuits.

The present implementation exhibits limitations due to pinning resulting in different temperature conditions used for optical switching and current-driven motion. Reducing pinning through stack optimization, using local optical or electrical heating, and designing tracks with controlled disorder could allow faster, more reproducible transport of optically programmed skyrmions. The essential advance established here is the new operation: deterministic optical control of the skyrmion core.

# 4 Experimental Section

## 4.1 Sample preparation

A double-wedge multilayer with the composition Ta(3)/Pt(5)/Ir(1)/CoB($t_{CoB}$)/Gd($t_{Gd}$)/Pt(2) was deposited by magnetron sputtering using a Singulus Rotaris system. The film was grown on a 4-inch Si/$SiO_2$ (100) wafer, with the CoB and Gd thicknesses varying along two orthogonal directions. The center-point thicknesses were 1.5 nm for CoB and 0.9 nm for Gd. The Pt/Ir/CoB interfaces provide perpendicular magnetic anisotropy and interfacial DMI, while the CoB/Gd bilayer provides the ferrimagnetic character [34] needed for all-optical helicity-independent switching.

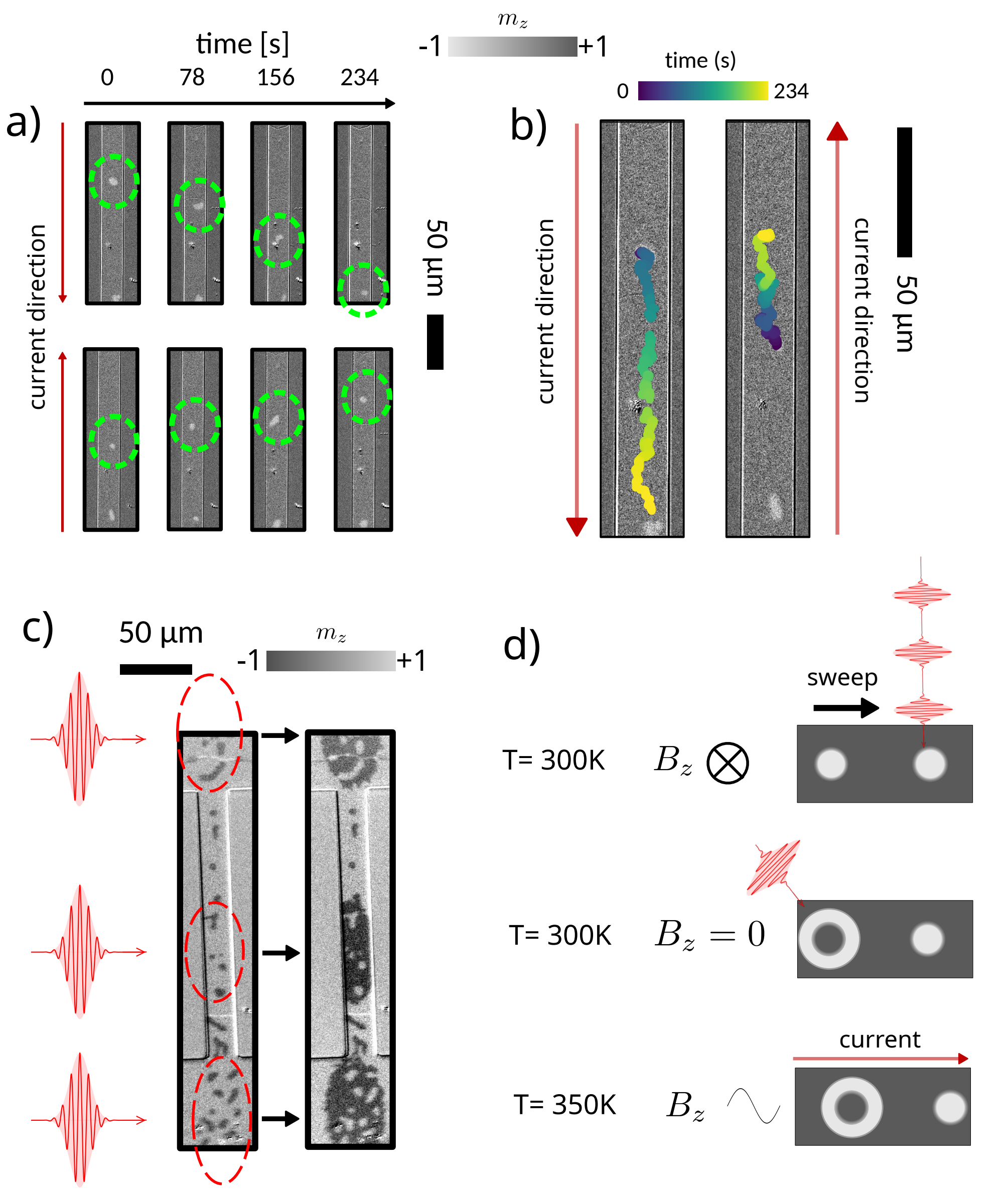

Figure 5: Current-driven transport in a device compatible with optical core switching. a) Kerr microscopy snapshots of skyrmion motion in a 20 $\mu$m wire at $T = 350$ K under $J_{\mathrm{DC}} = 5.9 \times 10^9$ A m$^{-2}$, $B_{\mathrm{AC}} = 1.380$ mT, $f_{\mathrm{AC}} = 50$ Hz, and $B_{\mathrm{DC}} = 0.06$ mT. The top and bottom image sequences correspond to opposite current polarities, showing reversal of the direction of motion. b) Skyrmion trajectories extracted from the experiments of (a). The red arrows indicate the direction of the current.c) Demonstration of AO-HIS in the same wire device at room temperature: skyrmions written by pulse-train scanning and selected textures switched by single 30 fs pulses. For the optical writing step, the laser parameters are $\tau = 30$ fs, $F = 1.95$ mJ cm$^{-2}$, and $f = 5$ kHz under $B_{\mathrm{DC}} = 0.15$ mT; for the switching step, $\tau = 30$ fs and $F = 2.92$ mJ cm$^{-2}$.d) Proposed schematic device workflow illustrating the sequence optical write $\rightarrow$ optical core switch $\rightarrow$ electrical transport.

After deposition, the wafer was diced into 1 cm × 1 cm pieces, corresponding to approximate CoB and Gd thickness steps of 0.1 nm. In the analysis, each piece was treated as compositionally uniform. Spatial maps of the coercivity were obtained from out-of-plane magneto-optical Kerr effect (MOKE) hysteresis loops to identify compositions combining skyrmion stability and optical switching. For current-driven experiments, a second sample was deposited using the same sputtering process on an insulating undoped $Si/SiO_2$ substrate and patterned into 20 $\mu$m × 187 $\mu$m wires by optical lithography using a Durham Magneto Optics Microwriter ML3 Pro.

### 4.2 Laser experiments

Laser experiments were performed using a home-built polar Kerr microscope coupled to an 800 nm ultrafast laser system. The setup enables single-pulse excitation with pulse durations ranging from 30 fs to 10 ps and tunable pulse energy. The fluence of single laser pulses was determined by fitting the sizes of switched domains as a function of pulse energy. Pulse-train operation was performed at a repetition rate of 5 kHz. In all experiments shown here, the laser polarization was linear.

The field of view of the Kerr microscope was 290 × 240 $\mu m^2$, with a spatial resolution of approximately 0.5 $\mu$m. An out-of-plane magnetic field was applied during selected experiments by placing an electromagnet close to the sample. This field was used to bias the magnetic textures nucleated during pulse-train laser writing.

### 4.3 Current injection experiments

Current-injection experiments were performed using a commercial evico Kerr microscope setup with a field of view of 260 × 160 $\mu m^2$ and a spatial resolution of approximately 300 nm. The setup allows for application of both out-of-plane and in-plane fields. The out-of-plane magnet was custom-made to allow field control with microtesla precision. The sample was mounted on a Peltier element positioned directly on top of the coil. The temperature was monitored with a Pt100 sensor placed next to the sample, providing a stability better than 0.1 K. The wire device was contacted using gold wire bonds on a custom printed circuit board, and a DC current was supplied using a Keithley2400 current source.

This system is characterized by some skyrmion pinning, as no visible skyrmion diffusion can be observed at 300 K [17]. This pinning poses a challenge to drive skyrmions efficiently with SOT, because the high current densities required for depinning produce a sizeable Oersted field, which can annihilate the textures. To reduce the effective pinning and enable motion at lower current densities, the temperature was increased to $T = 350$ K, and a small oscillating out-of-plane magnetic field was applied together with a small DC offset field. These two strategies have been shown to speed up skyrmion diffusion, corresponding to a reduced effective pinning [17,42] . At $T = 350$ K, the application of an oscillating (AC) OOP field of amplitude $B_{\mathrm{AC}} = 1.380$ mT and frequency $f_{\mathrm{AC}} = 50$ Hz, with constant offset $B_{\mathrm{DC}} = 0.06$ mT leads to visible skyrmion diffusion.

Skyrmions were nucleated under a constant out-of-plane field of 0.08 mT superimposed with the small AC field by applying an in-plane magnetic pulse of 40 mT. The constant field was then temporarily reduced to 0.048 mT to annihilate excess skyrmions and subsequently set to 0.06 mT to stabilize the remaining skyrmions. The trajectories of the skyrmions during these experiments were extracted by employing a machine-learning based skyrmion classifier tool [43], as shown in Figure 5b.

Our used temperature sequence allows for device functionality because the written and switched textures remain stable upon heating to 350 K, as noted in **Figure S4**.

**Declaration on the use of LLMs**

GPT6.5 Luna was used to format the text of the manuscript, in the final phases of the writing. No scientific content was altered by the use of LLMs in this work. The same LLM was also used to generate the Python code for the analysis and plotting of the data of Figure 5b, and to generate the Python code for the visualization of Figure 1d.

**Table of Contents**

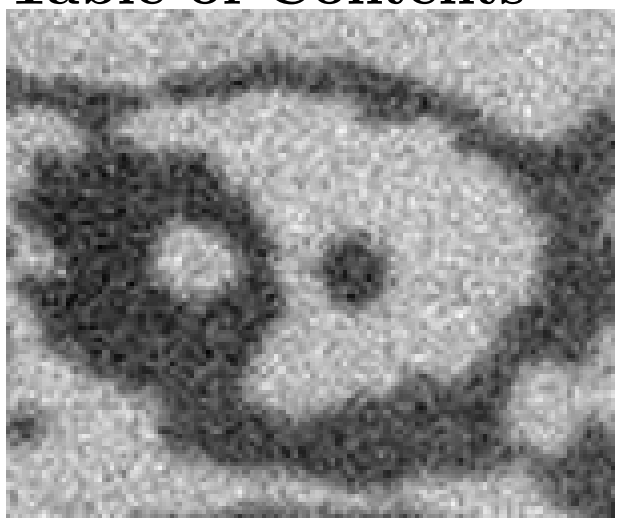

Skyrmions are magnetic microstructures with strong potential for low-power computing applications. Single laser pulses can invert these textures on ultrafast timescales. By combining this optical manipulation with existing electrical methods, new possibilities emerge for designing skyrmion-based devices.

**Acknowledgements**

This work is supported by the French National Research Agency (ANR) through the France 2030 government grants PEPR SPIN: Chirex (ANR-22-EXSP-0002) and SPINMAT (ANR-22-EXSP-0007) and Program "Lorraine Initiative of Excellence", (ANR-15-IDEX-04-LUE). This work is also supported by the Institute Carnot ICEEL, the Région Grand Est, the Metropole Grand Nancy for the project “OPTIMAG” and OPTI-MEMORY. This work has received funding from the European Union’s Horizon 2020 research and innovation program under the Marie Skłodowska-Curie Grant Agreement No. 860060 “Magnetism and the effect of Electric Field” (MagnEFi), as well as from ERC Synergy Grant No. 856538, project “3D-MAGiC,” the Horizon Europe Project No. 101070290 (NIMFEIA), and Horizon Europe Programme Horizon 1.2 under the Marie Skłodowska-Curie Actions (MSCA), Grant agreement No.101119608 (TOPOCOM). It has also been supported by the Deutsche Forschungsgemeinschaft (DFG, German Research Foundation) – SPP 2137 Skyrmionics (project 462597720), TRR 173 – 268565370 (projects A01, B02 and A12), and the Dynamics and Topology Center TopDyn funded by the State of Rhineland Palatinate. This work was partly supported by the National Research Council of Science & Technology (NST) grant by the Korean government, MSIT (Grant No. GTL24041-000). The authors would like to acknowledge José Omar Ledesma-Martín, Sachin Krishnia and Théo S. Balland for their support in the device design and data analysis.

# Single-pulse Optical Switching combined with Current-induced motion of Skyrmionic Spin Textures

Edoardo Mangini, Duc Minh Tran, Boonthum Kunyangyuen, Jun Xiao Lin, Raphael Gruber, Michel Hehn, Stephane Mangin*, and Mathias Kläui*

*E. Mangini, D. M. Tran, Dr. R. Gruber, Prof. M. Klaui*
Institute of Physics, Johannes Gutenberg University Mainz, Staudingerweg 7, 55128 Mainz, Germany
Email: Klaeui@uni-mainz.de
*B. Kunyangyuen, Dr. J. X. Lin, Prof. M. Hehn, Prof. S. Mangin*

Email: stephane.mangin@univ-lorraine.fr
Institut Jean Lamour, Universite de Lorraine, CNRS, Campus Artem, 54000 Nancy, France

*Prof. M. Hehn, Prof. S. Mangin*

Center for Science and Innovation in Spintronics, Tohoku University, Sendai, Japan

# Figure S1

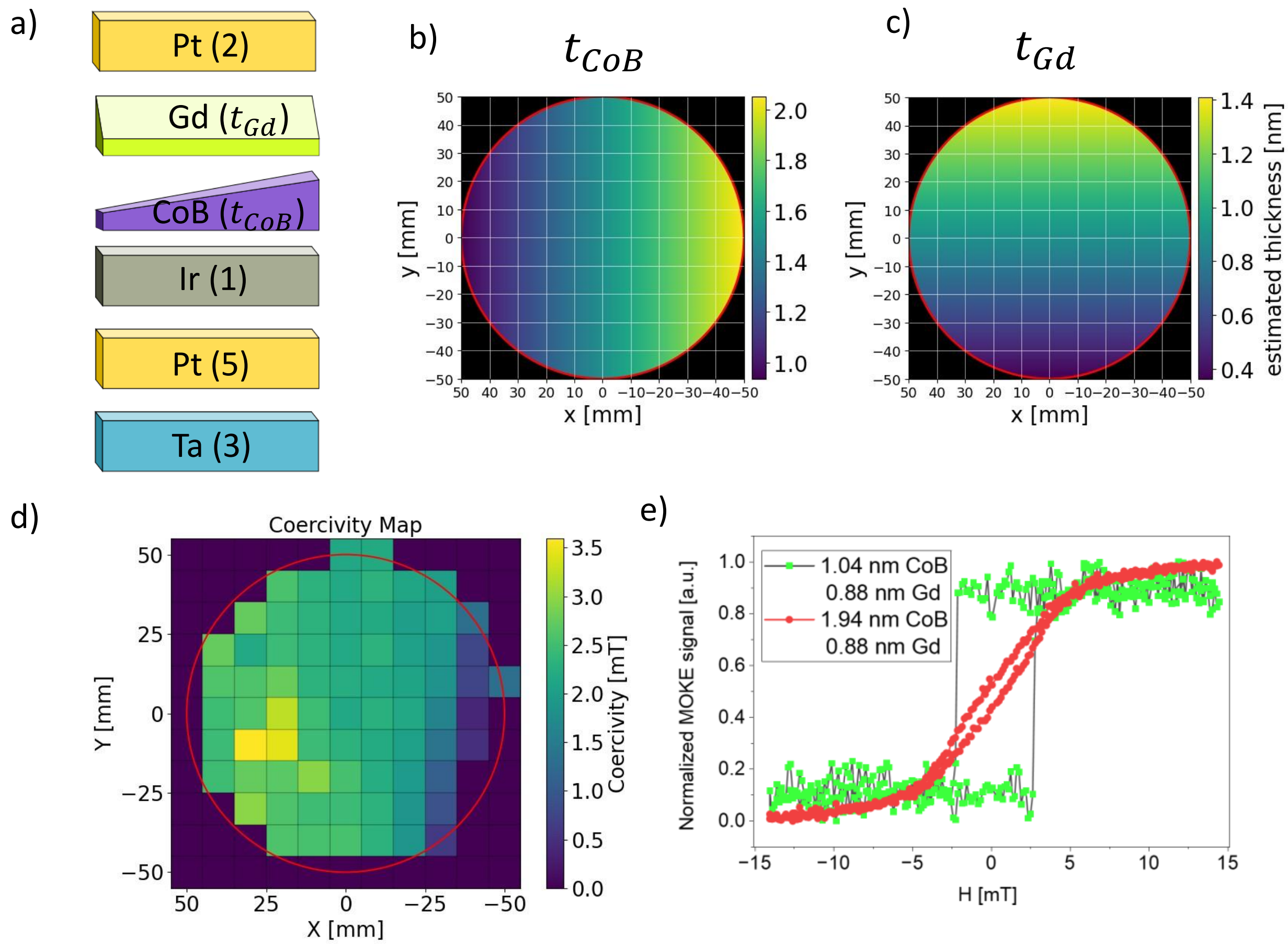

***Figure S1.*** *Characterization of the double-wedge sample for skyrmion all-optical switching. a) Schematic of the stack. b,c) Estimated thickness distributions for the CoB and Gd wedges. d) Map of the measured coercivity obtained by MOKE. e) Representative hysteresis loops measured at positions corresponding to different CoB/Gd compositions.*

Figure S1 shows the material stack used for skyrmion all-optical switching. The sample was sputtered on a 4-inch Si/$SiO_2$ substrate, and has a double wedge structure (see fig. S1a). Estimations of the thickness gradients are shown in Fig. S1b-c. The magnetic properties of the sample depend heavily on the CoB/Gd composition, as exemplified by the magnetic coercivity map shown in Figure S1d. Fig. S1e shows two representative hysteresis loops corresponding to two different CoB thicknesses.

# Figure S2

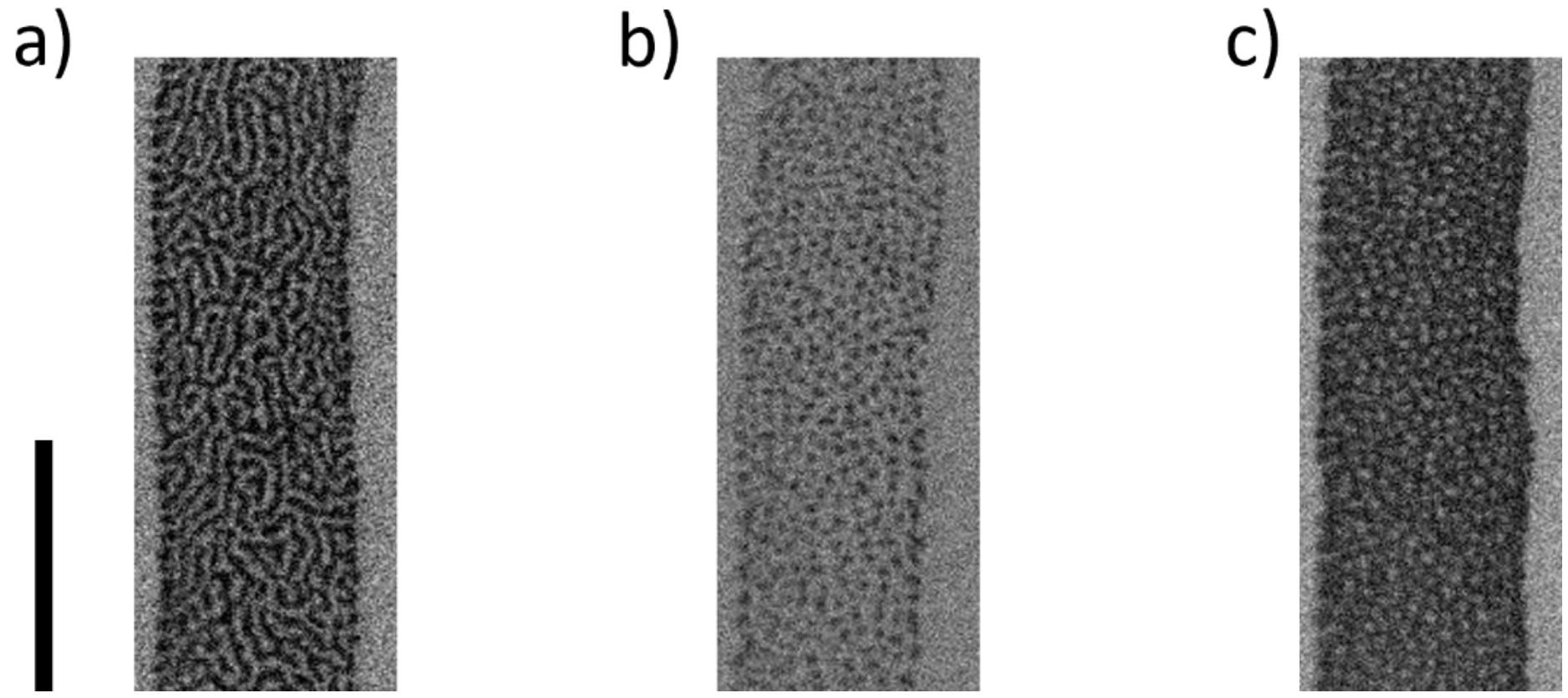

*Figure S1 Effect of the out-of-plane magnetic field during laser writing. MOKE microscopy images of magnetic textures written in pulse-train mode under out-of-plane fields of a) +0.2 mT, b) +0.7 mT and c) -0.7 mT. By changing the applied field during the scan, the written state can be tuned from stripe domains to skyrmions of opposite contrast. Scale bar: 50 µm.* Laser parameters: $f = 5\ kHz, F = 6.19\ mJ/cm^2, \tau = 1\ ps$. Magnetic thicknesses: $t_{CoB} = 1.66\ nm, t_{Gd} = 0.62\ nm$.

Images taken at RT.

Figure S2 shows the effect of changing the magnetic field on the skyrmion writing step. By changing the applied OOP field, either stripes (Fig S2a), Black skyrmions (Fig S2b) or White skyrmions (Figure S2c) can be nucleated. This behaviour is consistent with earlier reports showing that optical excitation can be used to nucleate skyrmions stochastically or deterministically.[1] After the laser spot is moved away, the applied field can be changed in a wider range (while still below the coercivity of the material), without influencing the magnetic structures

# Figure S3

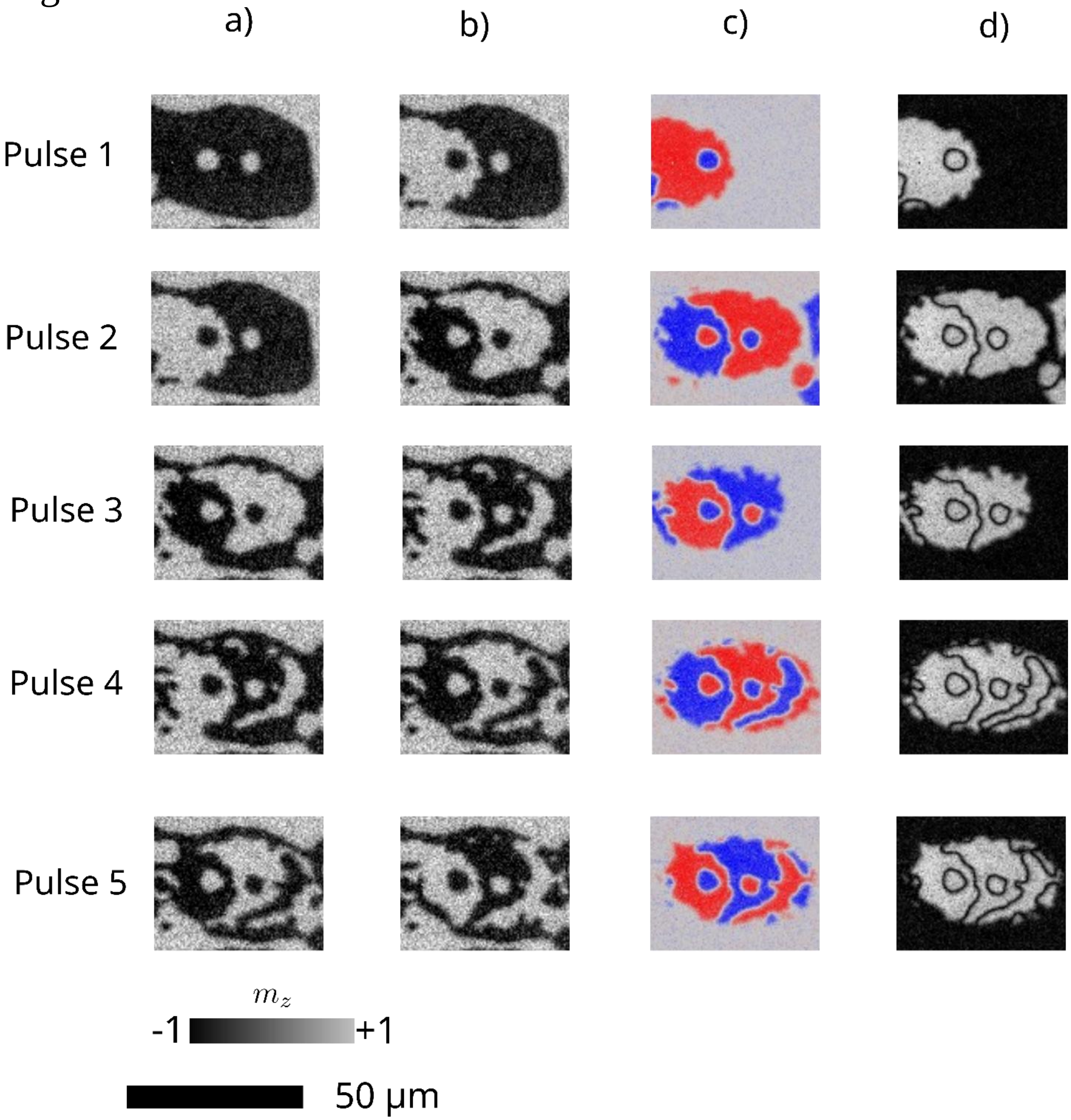

*Figure S3 Image analysis of 5 pulse all-optical switching sequence. a) initial state (before the pulse); b) (final state (after pulse) ;c) difference state (subtracting image a) from image b)); d) absolute value of contrast shown in image c).*

Figure S3 shows the 5- pulse all optical switching sequence. After each of the shots, an area encompassing one or more skyrmions is switched. Figure S3 c shows the difference image between initial and final state. Only the elliptical area irradiated via ultrafast laser pulse undergoes switching. Figure S3 d shows the plots of the absolute values of the contrast of the figures in S3c. In this column, a black contrast corresponds to no change in magnetic contrast before and after the pulse. Interestingly, the domain walls of the skyrmions correspond to sharp black lines, indicating that no skyrmion motion is triggered by the laser pulse.

# Figure S4

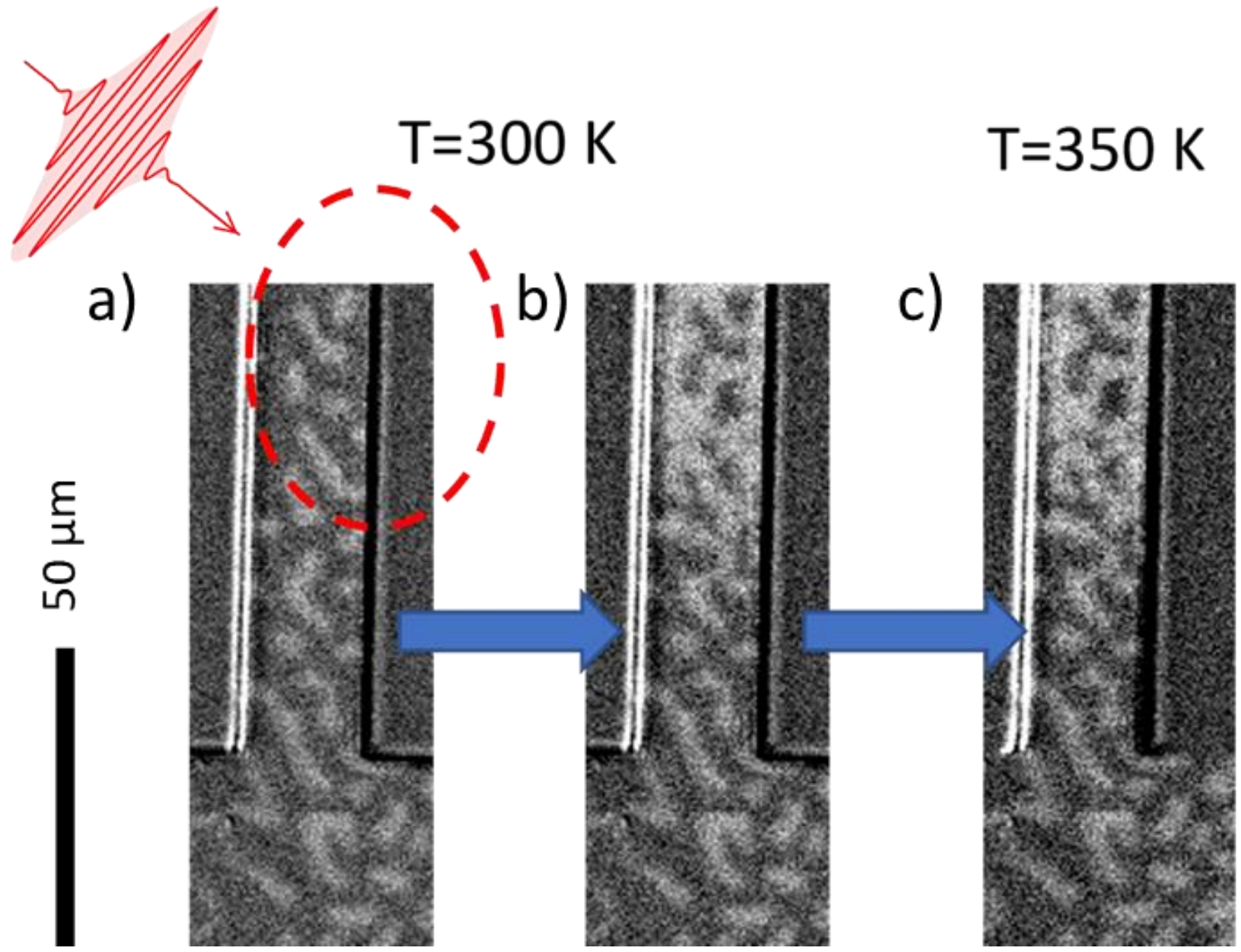

*Figure S4 Thermal stability of written and optically switched skyrmions. a) Skyrmions nucleated by laser scanning at room temperature. b) Image after inducing AOHIS with a single laser pulse in the indicated region. c) Image after heating the sample to 350 K. The magnetic textures remain stable upon heating from room temperature to 350 K, supporting the proposed sequence of optical programming followed by current-driven transport. Scale bar: 50 µm*

Figure S4 shows the stability of the nucleated and switched skyrmions with temperature. The magnetic features do not show any evident modification when heating the sample from room temperature (center panel) to 350 K (right panel).